\documentclass[
    ,final            
  ]
  {aipproc}

\layoutstyle{6x9}

\begin{document}

\title{Similarities in the temporal properties of gamma-ray bursts and soft gamma-ray repeaters}

\author{S. McBreen, L. Moran, B. McBreen, L. Hanlon,\\ J. French, M. Conway}{
  address={Department of Experimental Physics, University College Dublin, Dublin 4, Ireland.}
}

\begin{abstract}
Magnetars are modelled as sources that derive their output from magnetic
energy that substantially exceeds their rotational energy \cite{Thompson:1992}. 
An implication of the  recent polarization measurement of
GRB 021206 is that the emission mechanism may be dominated by a magnetic field 
that originates in 
the central engine \cite{Coburn2003}. Similarities in the temporal properties of
 SGRs and GRBs are considered
in light of the fact that the central engine in GRBs may be magnetically 
dominated. The results show that 1) the time intervals
between outbursts in SRG 1806\,$-$\,20 and pulses in GRBs are consistent
with lognormal distributions and 2) the cumulative outputs
of SGRs and GRBs increase linearly with time. 
This behaviour can be successfully modelled by a relaxation system
that maintains a steady state situation.

\end{abstract}

\maketitle


\section{Introduction}
Over the past decade the evidence for neutron stars with ultra-strong
magnetic fields or magnetars has become convincing. Soft gamma-ray
repeaters (SGRs) are associated with supernova remnants and 
have multiple bursts of gamma-rays which distinguish them from gamma-ray
bursts (GRBs) e.g. \cite{Zhang:2003,DeRujula:2003}. SGRs have intensely active periods which 
can last for weeks or months that are separated by quiescent 
phases lasting for years or decades \cite{Mazets:1979}. The most intense outburst
recorded to date has $\sim$\,10$^{44}$ ergs in $\gamma$-rays.
Several SGRs and anomalous X-ray pulsars have been found to be
X-ray pulsars that have unusually high spin down rates. The rapid
reduction in spin is usually attributed to magnetic breaking caused by the super-strong
magnetic fields with values above 10\,$^{14}$ G. In the magnetar model the magnetic field
provides the burst energy \cite{Thompson:1992}. A common scenario is that stresses build up
in the magnetic field and cause a quake in the crust of the neutron star
which ejects plasma Alfv\`en waves through the magnetosphere.

GRBs are 
non-recurrent catastrophic events that radiate $\sim$\,10$^{51}$ ergs in
$\gamma$-rays. There is substantial evidence that GRBs are linked to 
supernova explosions e.g. \cite{Hjorth:2003}. 
They are often modelled as accretion onto 
a newly formed black hole e.g. \cite{macfad:1999}. Recent measurements suggest that
the $\gamma$-rays are highly polarised 
by the magnetic field which may originate in
the central engine with values of 
B above 10$^{17}$ G  \cite{Coburn2003}.
It is therefore interesting to compare the pulse properties of SGRs and
GRBs. 
It has been noted that the lognormal distributions apply to the pulse properties of SGRs 
\cite{hmrs:1994,gogus:1999,gwk:2000} and GRBs \cite{mhlm:1994,lifen:1996,sheila:2001}.
Furthermore the radio afterglow from
the giant flare from SGR 1900 $+$ 14 is similar to radio afterglows from GRBs
\cite{Cheng:2003}.

\section{Time intervals between outbursts in SGRs and GRBs}
\begin{figure}
\label{time_int_sgr}
  \includegraphics[width=.43\textwidth]{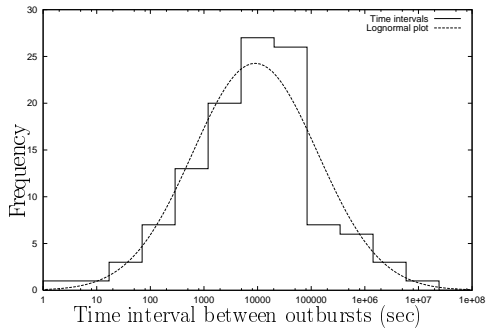}
\vspace{1cm}\\
\caption{
 Histogram of the number of bursts from SGR 1806-20 versus time
      interval between the bursts. The dashed curve is a lognormal
      fit to the data.
The SGR was observed by the International Cometary Explorer (ICE) 
and the data are taken from Ulmer et al \cite*{Ulmer:1993}.}
\end{figure}

\begin{figure}
\label{time_int_grb}
  \includegraphics[width=.9\textwidth]{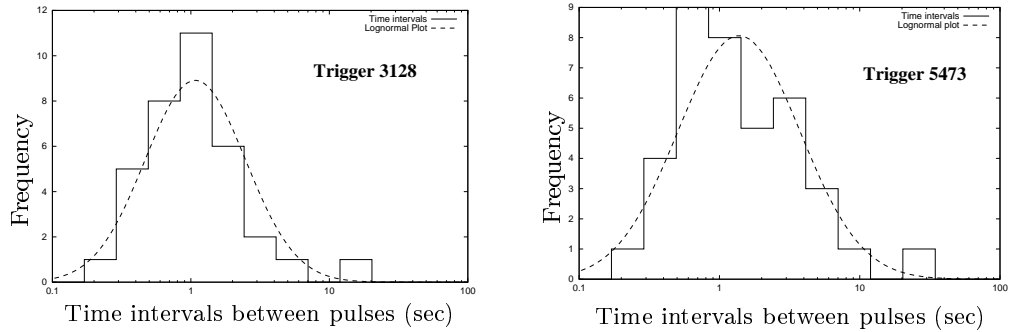} \\
\caption{  Histograms of the time interval between the pulses
 and lognormal representations for BATSE triggers GRB 940817
 and GRB 960524
 with 37 and 39 pulses respectively.}
 \vspace{-1em}
\end{figure}

SGR 1806 $-$20 had a very active phase during 1983 when more than
100 outbursts were recorded \cite{Laros:1987,Ulmer:1993,hmrs:1994}. The distribution of the 
time intervals between the outbursts is given in Figure \ref{time_int_sgr}
along with a lognormal fit to the data. Long time intervals between
active phases ($>$ 10 years) can easily be accomodated by the long 
tail of the lognormal distribution \cite{hurley:1995}.
The time intervals between pulses in two 
BATSE GRBs with large numbers of pulses 
are given in Figure \ref{time_int_grb} along with lognormal plots.
The pulse properties and time intervals between
the pulses in short and long
GRBs have been found to be lognormally
distributed \cite{mhlm:1994,lifen:1996,gupta:2000,sheila:2001,quilligan:2002,nakar_prian:2002}.
It is interesting to note that the time intervals between radio
glitches in the Vela pulsar are lognormally distributed \cite{hmrs:1994}.

In Fig. 1 the median time interval between SGR outbursts is $\sim$ 10$^{4}$\,s and 
in Fig. 2
the time interval between pulses in the GRBs is $\sim$ 1\,s.

\vspace{-0.5em}
\section{Cumulative light curves for SGRs and GRBs}
\begin{figure}
\label{sgr_cumul}
 \includegraphics[width=.43\textwidth]{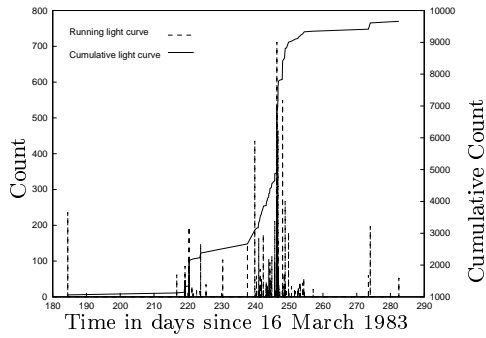}
\vspace{1cm}
\caption{ The running (dashed line) and cumulative (solid line)
lightcurves of the outbursts from SGR 1806 $-$ 20 in 1983. The presentation is adapted from 
Palmer \cite*{palmer:1999}. The cumulative output approximates to straight
lines of different slopes in the active region
in late 1983. } 
\end{figure}

\begin{figure}
\label{grb_cumul}
 \includegraphics[width=.9\textwidth]{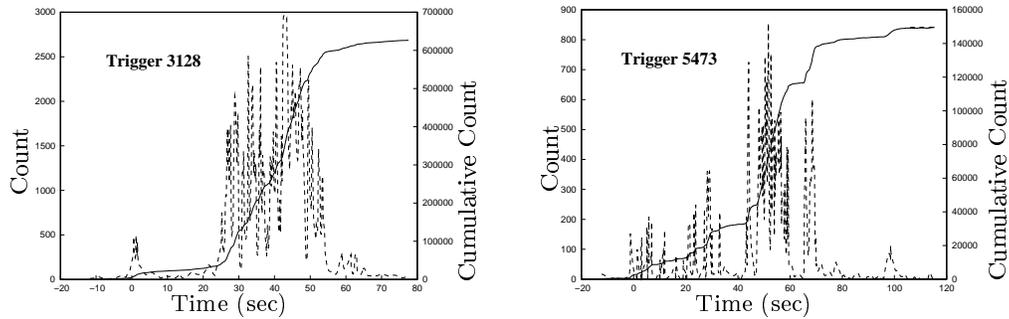} \\
\caption{ The running light curves (dashed line) and cumulative light 
curves (solid line) of GRB 940817 and GRB 960524
The cumulative light curves are 
approximately linear over the majority of the burst. The cumulative
light curve of trigger 5473 can be fit by two linear sections.
Trigger 3128 can be fit by one straight line despite the temporal complexity.}
\vspace{-3em}
\end{figure}

The cumulative light curve has been shown to be an interesting parameter for 
SGRs \cite{palmer:1999} and GRBs \cite{mcbreenb:2002}. The running and cumulative light curves for the outbursts 
from SGR 1806 $-$ 20 are given in Figure \ref{sgr_cumul}. 
The cumulative light curves of a large majority of short and long
GRBs has been found to be approximately linear with time. 
The running and cumulative light curves of the two GRBs in 
Figure \ref{time_int_grb} 
are given in Figure \ref{grb_cumul}. The cumulative output increases linearly with time over the most active 
part of the GRB. 

The cumulative output of the large majority of long and short GRBs was
found to increase linearly with time \cite{mcbreenb:2002}. 

\vspace{-1em}
\section{Relaxation systems}
The behaviour of bursting sources can be
usefully compared with relaxation
systems \cite{palmer:1999,mcbreenb:2002}. A relaxation system is
 taken to be one that
continuously accumulates energy from some process and
discontinuously releases it. The energy in the reservoir at any
time $t$ is
\begin{equation}
E(t) = E_{o} + \int_{o}^{t} R(t)dt - \Sigma S_{i}
\end{equation}
where $E_{o}$ is the energy stored in the reservoir that
accumulates energy at a rate $R(t)$ and discontinuously releases
events of size $S_{i}$.

The simplest system is referred to as a relaxation oscillator
where there is a fixed level or trip-point that triggers a
release of the energy when E = E$_{\rm max}$.  The SGR
and GRB pulses are not
consistent with this type of oscillator.
 More complicated behaviour
occurs when the accumulation rate, trigger rate or release
strength are not constant \cite{rm:2001}.
If the system starts from a minimum
level E = E$_{\rm min}$, accumulates energy at a constant rate R
= r, the sum of the releases is approximately a linear function
of time i.e. \(\Sigma S_{i} \propto rt\).  This model can account
for the approximately linear increase in cumulative counts.
 The output from GRBs and SGRs has a tendency to keep the cumulative
count close to a linear function and maintain a steady state situation

In a minority of bright GRBs with many pulses it was found that the
cumulative output increases with time as t$^{2}$ in $\sim$ 10\% of bursts and as (1-t)$^{2}$ in
$\sim$5\%. This behaviour was attributed to  
a change in spin of the newly formed black hole \cite{mcbreenS:2002}.


\vspace{-1.25em}
\section{Acknowledgement}
SMcB acknowledges IRCSET grant number RS/2002/820-8M for support. 
LH, BMcB and JF
thank IRCSET Basic Research Grant number SC-2002-377 for support. 
\vspace{-1.25em}





\begin{thebibliography}{10}
\providecommand{\enquote}[1]{``#1''}
\expandafter\ifx\csname url\endcsname\relax
  \def\url#1{\texttt{#1}}\fi
\expandafter\ifx\csname urlprefix\endcsname\relax\def\urlprefix{URL }\fi

\bibitem{Thompson:1992}
{Duncan}, R., and {Thompson}, C., \emph{ApJ}, \textbf{392} (1992).

\bibitem{Coburn2003}
{Coburn}, W., and {Boggs}, S.~E., \emph{Nature}, \textbf{423}, 415--417 (2003).

\bibitem{Zhang:2003}
{Zhang}, B., and {Meszaros}, P. (2003), [astro-ph/0311321].

\bibitem{DeRujula:2003}
{Dar}, A., and {De Rujula}, A. (2003), [astro-ph/0308248].

\bibitem{Mazets:1979}
{Mazets}, E.~P., {Golentskii}, S.~V., {Ilinskii}, V.~N., {Aptekar}, R.~L., and
  {Guryan}, I.~A., \emph{Nature}, \textbf{282}, 587--589 (1979).

\bibitem{Hjorth:2003}
{Hjorth}, J. et al.
  , \emph{Nature}, \textbf{423}, 847--850 (2003).

\bibitem{macfad:1999}
{MacFadyen}, A.~I., and {Woosley}, S.~E., \emph{ApJ}, \textbf{524}, 262--289
  (1999).

\bibitem{hmrs:1994}
{Hurley}, K.~J., {McBreen}, B., {Rabbette}, M., and {Steel}, S., \emph{A\&A},
  \textbf{288}, L49--L52 (1994).

\bibitem{gogus:1999}
{G\"o\^g\"us}, E., {Woods}, P.~M., {Kouveliotou}, C., {van Paradijs}, J.,
  {Briggs}, M.~S., {Duncan}, R.~C., and {Thompson}, C., \emph{ApJ},
  \textbf{526}, L93--L96 (1999).

\bibitem{gwk:2000}
{G\"{o}g\"{u}s}, E., {Woods}, P.~M., {Kouveliotou}, C., {van Paradijs}, J.,
  {Briggs}, M.~S., {Duncan}, R.~C., and {Thompson}, C., \emph{ApJ},
  \textbf{532}, L121--L124 (2000).

\bibitem{mhlm:1994}
{McBreen}, B., {Hurley}, K.~J., {Long}, R., and {Metcalfe}, L., \emph{MNRAS},
  \textbf{271}, 662 (1994).

\bibitem{lifen:1996}
{Li}, H., and {Fenimore}, E.~E., \emph{ApJ}, \textbf{469}, L115 (1996).

\bibitem{sheila:2001}
{McBreen}, S., {Quilligan}, F., {McBreen}, B., {Hanlon}, L., and {Watson}, D.,
  \emph{A\&A}, \textbf{380}, L31 (2001).

\bibitem{Cheng:2003}
{Cheng}, K.~S., and {Wang}, X.~Y., \emph{ApJ}, \textbf{593}, L85--L88 (2003).

\bibitem{Laros:1987}
{Laros}, J.~G. et al.,
  \emph{ApJ}, \textbf{320}, L111 (1987).

\bibitem{Ulmer:1993}
{Ulmer}, A., {Fenimore}, E.~E., {Epstein}, R.~I., {Ho}, C., {Klebesadel},
  R.~W., {Laros}, J.~G., and {Delgado}, F., \emph{ApJ}, \textbf{418}, 395
  (1993).

\bibitem{hurley:1995}
{Hurley}, K.~J., {McBreen}, B., {Delaney}, M., and {Britton}, A.,
  \emph{Ap\&SS}, \textbf{231}, 81--84 (1995).

\bibitem{gupta:2000}
{Gupta}, V., {Das Gupta}, P., and {Bhat}, P.~N., \enquote{AIP Conf. Proc. 526:
  Gamma-ray Bursts, 5th Huntsville Symposium. Edited by R.M. Kippen et al.,}
  2000, p. 215.

\bibitem{quilligan:2002}
{Quilligan}, F., {McBreen}, B., {Hanlon}, L., {McBreen}, S., {Hurley}, K.~J.,
  and {Watson}, D., \emph{A\&A}, \textbf{385}, 377--398 (2002).

\bibitem{nakar_prian:2002}
{Nakar}, E., and {Piran}, T., \emph{MNRAS}, \textbf{331}, 40--44 (2002).

\bibitem{palmer:1999}
{Palmer}, D.~M., \emph{ApJ}, \textbf{512}, L113--L116 (1999).

\bibitem{mcbreenb:2002}
{McBreen}, S., {McBreen}, B., {Hanlon}, L., and {Quilligan}, F., \emph{A\&A},
  \textbf{393}, L29--L32 (2002).

\bibitem{rm:2001}
{Ramirez-Ruiz}, E., and {Merloni}, A., \emph{MNRAS}, \textbf{320}, L25 (2001).

\bibitem{mcbreenS:2002}
{McBreen}, S., {McBreen}, B., {Hanlon}, L., and {Quilligan}, F., \emph{A\&A},
  \textbf{393}, L15--L19 (2002).

\end{thebibliography}

\IfFileExists{\jobname.bbl}{}
 {\typeout{}
  \typeout{******************************************}
  \typeout{** Please run "bibtex \jobname" to optain}
  \typeout{** the bibliography and then re-run LaTeX}
  \typeout{** twice to fix the references!}
  \typeout{******************************************}
  \typeout{}
 }

\end{document}